\documentclass[11pt]{article}
\usepackage[preprint]{acl}
\usepackage{times}
\usepackage{latexsym}
\usepackage[T1]{fontenc}
\usepackage[utf8]{inputenc}
\usepackage{microtype}
\usepackage{inconsolata}
\usepackage{graphicx}
\usepackage{amsmath,amssymb,amsfonts}
\usepackage{colortbl}
\definecolor{blkSFT}{HTML}{F4F4F4}
\definecolor{blkText}{HTML}{EAF1FA}
\definecolor{blkOnline}{HTML}{FCF2E8}
\definecolor{blkRAG}{HTML}{EBF6EC}
\definecolor{blkTool}{HTML}{F2ECFA}
\definecolor{rowRepro}{HTML}{E4E4E4}
\definecolor{rowOurs}{HTML}{D6E8FB}
\usepackage{booktabs} % Recommended for better-looking tables
\usepackage{tabularx}
\usepackage{array}
\usepackage{multirow}
\usepackage[most]{tcolorbox}
\definecolor{trajSys}{HTML}{1A5276}
\definecolor{trajUsr}{HTML}{1E8449}
\definecolor{trajAsst}{HTML}{8C5A0A}
\definecolor{trajTool}{HTML}{6C3483}
\definecolor{trajBg}{HTML}{FAFBFC}
\definecolor{trajFrame}{HTML}{B0BEC5}

\newcolumntype{Y}{>{\raggedright\arraybackslash}X}

\newcommand{\protagent}{ProtAgent}
\newcommand{\tgsvr}{TGSVR}

\title{ProtAgent: Post-Training Compact Protein Tool Agents with Interaction-Level Credit Assignment}
\author{
  Yuxiang Wang, Baojian Zhou, Qifang Zhao,\\
  Keyue Jiang, Yanghua Xiao, and Xiaoxiao Xu
}
\date{}
\begin{document}

\maketitle
\begin{abstract}
Protein annotation agents must choose suitable bioinformatics tools, interpret their outputs,
and use the resulting evidence in the final answer. Most existing systems rely on frontier-scale
models. We ask whether a compact model can instead learn this behavior through post-training
with real tool interaction. We present \protagent{}, a two-stage pipeline that uses supervised
fine-tuning (SFT) to learn the tool interface and standard on-policy REINFORCE to improve the
interaction policy. Intermediate tool calls receive action rewards, and the final response
receives a terminal answer reward. Step-level returns assign these rewards to the responses that
produced them and to preceding actions. The episode-level baseline instead assigns one episode return
to every action. On a fixed closed-loop protocol
with real bioinformatics tools, \protagent{}-8B achieves 61.00\% success under the
annotation-coverage metric, compared with 41.60\% for DeepSeek-V3.2-Thinking 671B and 60.80\%
for the strongest evaluated frontier system. In the three-seed comparison, mean success is
42.70\% with episode-level returns and 61.40\% with step-level returns under identity return
normalization; the corresponding group-relative results are 36.50\% and 58.20\%.
The metric measures recovery of reference annotations
under this protocol; it does not by itself establish biological correctness or complete evidence
grounding.
\end{abstract}

\section{Introduction}

Protein function annotation is an evidence-integration problem rather than a direct reading of
sequence. Community assessments such as CAFA show that automated annotation remains challenging
and benefits from combining heterogeneous evidence~\citep{radivojac2013cafa}. Sequence-similarity
search connects a query to characterized homologs~\citep{mmseqs2}; profile models and integrative
resources identify families, domains, and functional sites~\citep{pfam,interpro2022}; specialized
predictors expose signal peptides and disordered regions~\citep{signalp6,iupred3}; and curated
databases provide reviewed records and controlled ontology terms~\citep{uniprot,geneontology,
disprot2024}. A tool-using agent must therefore decide which evidence is needed, invoke suitable
tools, interpret heterogeneous observations, and stop with an answer that satisfies the question.
We evaluate this workflow through protein question answering. Each question specifies an
annotation target, and the final answer is scored against the corresponding reference annotation.

The PFUA system shows that access to bioinformatics tools can improve protein function analysis
with large language models~\citep{fan2026pfua}. Its strongest systems use online frontier models,
while its smaller post-trained model plateaued when trained only with a final-answer overlap
reward. This leaves a different question open: can a compact model learn how to gather and use
tool evidence when it receives feedback during the interaction? Supervised examples can teach
tool syntax and basic routing, but they do not directly optimize which evidence to collect or
when to stop. A single terminal reward also gives the same credit to useful, redundant, and
malformed calls. We therefore study whether feedback at individual interaction steps improves
annotation recovery.

\protagent{} has two post-training stages. Supervised fine-tuning (SFT) on validated tool
trajectories teaches tool-call syntax, task-to-tool routing, and answer format. Reinforcement
learning (RL) then lets the model interact on-policy with the same tools. We use standard
critic-free REINFORCE; the main design choice is where reward is assigned. Intermediate calls
receive local feedback, the final answer receives a terminal reward, and Monte Carlo returns
carry these signals back through the same trajectory. A completed-tool guard prevents the model
from earning format reward without executing a tool.
We call each assistant response boundary---a tool call or the final answer---an interaction
step, and use \emph{interaction-level} and \emph{step-level} for this same return granularity.

We train separate models for two evaluation protocols. They share the Qwen3-8B base
model~\citep{yang2025qwen3}, but
use different training data, rewards, checkpoints, prompts, and test sets. The closed-loop
protocol is our main controlled experiment and covers five protein annotation tasks. The second
protocol follows PFUA for comparison with prior work. We report the two sets of results
separately. On the 1,000-example closed-loop test, \protagent{}-8B reaches 61.00\% success,
compared with 41.60\% for DeepSeek-V3.2-Thinking 671B and 60.80\% for the strongest evaluated
frontier system. Thus, under this protocol, an 8B local model is competitive with the evaluated
frontier-scale systems while remaining deployable without a hosted frontier model. These results
are specific to the stated scorer and tool setup.

\paragraph{Contributions.}
We make three contributions:
\begin{enumerate}
  \item We build a post-training pipeline for a compact protein agent that learns from real
  bioinformatics tool interactions, with SFT for the interface and RL for tool use and stopping.
  \item We compare step-level and episode-level returns in a three-seed $2\times2$ study that
  also varies return normalization, and find a large, consistent advantage for step-level
  returns under both normalization choices.
  \item We show that a local 8B model reaches 61.00\% success under a fixed closed-loop
  protocol, 19.40 points above the evaluated 671B baseline and competitive with the strongest
  evaluated frontier system. We analyze its gains and errors and report a separate
  PFUA-compatible evaluation.
\end{enumerate}

\begin{figure*}[t]
\centering
\includegraphics[width=\textwidth]{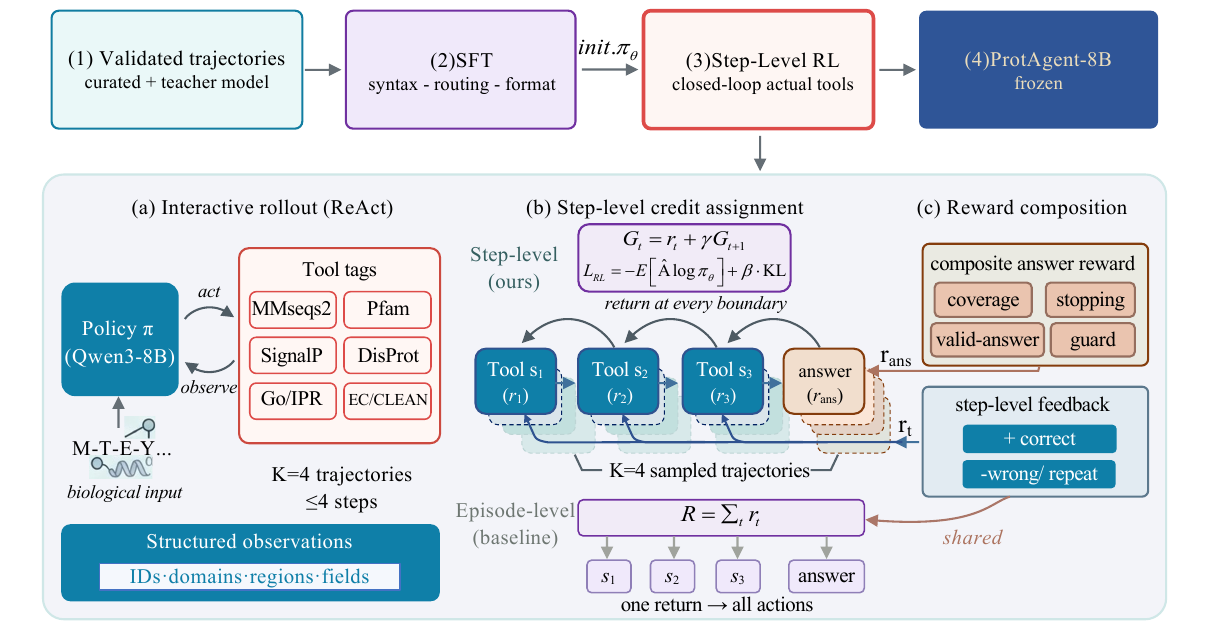}
\caption{\protagent{} overview. SFT learns the tool-use interface from validated trajectories;
RL then improves the closed-loop interaction policy. The three warm-restart stages are part of
the RL schedule, not a third post-training phase. During RL, the policy samples $K{=}4$
tool-use trajectories with at most four interaction steps. Batched Step-Level REINFORCE assigns
$G_t=r_t+\gamma G_{t+1}$ at each response boundary, whereas the episode-level baseline assigns
$R=\sum_t r_t$ to all actions. Intermediate tool calls receive action rewards and the final
answer receives the terminal answer reward; Appendix~\ref{app:tgsvr} gives the full reward
definition.}
\label{fig:overview}
\end{figure*}

\section{Related Work}
\paragraph{Protein tool-use agents.}
Protein-function agents combine language models with sequence-property, homology, domain, and
topology tools. PFUA demonstrates that inference-time access to these tools can improve protein
function understanding, and its strongest systems prompt online models~\citep{fan2026pfua}.
Companion agents refine Gene Ontology predictions by iteratively querying annotation
evidence~\citep{zhapa2026goagent}; multi-agent variants extend this approach to broader protein analysis and design
workflows~\citep{huang2025protchat,ghafarollahi2024protagents}. BioAgent Bench broadens
bioinformatics-agent evaluation to end-to-end analysis pipelines and motivates inspecting
intermediate work, not only final outputs~\citep{fa2026bioagentbench}. These systems primarily
evaluate frozen agents at inference time. A recent protein-set annotation benchmark likewise
examines frozen general-purpose agents and documents evidence-integration failures on complex
protein families~\citep{zhang2026proteinset}. \protagent{} instead studies how to post-train a
compact local model in a protein tool environment, with feedback assigned to individual
interactions.

\paragraph{Rewards and credit assignment for tool-using agents.}
ReAct established the alternating reasoning--action pattern used by many
tool-augmented agents~\citep{react}. One line of work assigns local rewards to tool selection,
argument use, or plan completion~\citep{qian2025toolrl,li2025rltr}. A second line changes how
credit is distributed across interaction boundaries. MT-GRPO, GiGPO, GTPO, and MatchTIR define
turn- or state-level advantages through group comparisons or trace supervision
\citep{wei2025mtgrpo,gigpo,ding2026gtpo,qu2026matchtir}. SLEA-RL combines step-level credit with
experience retrieval~\citep{wang2026slea}; TRACE derives turn rewards from reference-model state
values~\citep{tao2026trace}; and CrEST combines verifier-anchored turn advantages with
self-teacher token weighting~\citep{wang2026crest}. Learned critics and process reward models
provide another source of dense supervision for search and scientific reasoning
\citep{zhang2026criticsearch,zhao2026sciprm}. Other systems learn tool use mainly from outcome
feedback in code, search, or general tool environments
\citep{retool,searchr1,torl,toolzero}.
More recent methods derive step credit from answer-backtracked clues, redistribute dynamic-rubric
judgments to the responsible steps, adaptively mix step- and episode-level advantages using policy
uncertainty, or estimate hindsight success potentials from pooled transition graphs
\citep{lu2026abseeker,gandhi2026draco,liang2026gaca,zhu2026graphhca}.

\protagent{} instead uses fixed environment rewards and standard Monte Carlo returns at assistant
response boundaries. Our contribution is a controlled test of this assignment in a protein-tool
environment, not a new general-purpose credit-assignment estimator.

\paragraph{Evidence faithfulness in retrieval-augmented reasoning.}
VERITAS introduces turn-level faithfulness rewards for agentic search, including whether an
answer is entailed by retrieved information~\citep{xu2025veritas}, and AutoRefine combines
retrieval-specific rewards with iterative evidence refinement~\citep{shi2025autorefine}. Our
optional trajectory-fidelity reward applies that principle by comparing final answers with
observations from the same trajectory.

\section{Protein Tool-Use Setting and Protocols}
\label{sec:protocol}
\subsection{Task and Interaction Environment}
Each example contains a protein sequence $x$ and a natural-language question $q$ that specifies
an annotation target, such as an enzyme class, domain, signal peptide, disordered region, or
GO/InterPro record. The question is therefore an evaluation interface for protein annotation,
not an unrestricted biomedical question-answering task. At interaction step $t$, the policy emits
an action $a_t$: either a structured tool call or a terminal answer block. For a tool call, the
environment executes the requested bioinformatics program or database lookup and returns a
structured observation $o_t$. The interaction history is
$h_t=(x,q,a_1,o_1,\ldots,a_{t-1},o_{t-1})$, and the episode ends when the policy emits answer
$y$ or exhausts the horizon $H$.

An \emph{actual-tool trajectory} records outputs returned by the configured executables and
databases, not tool-like text generated by a model or a placeholder. Let
$O_\tau=\{o_1,\ldots,o_m\}$ denote the observations actually obtained in trajectory $\tau$.
The policy must decide what evidence to collect, when to stop, and how to convert $(q,O_\tau)$
into a task-compliant answer. We refer to the last step as \emph{evidence conversion}. The main
metric measures recovery of the reference annotation. We report observation-support measures
separately as behavioral diagnostics.

\subsection{Two Independent Evaluation Protocols}
We use two protocols and train a separate model for each. The models share a Qwen3-8B-Instruct
base, but not SFT data, RL rewards, checkpoints, prompts, or evaluation sets. The
\emph{closed-loop} protocol is the main experiment. It uses a balanced 100-example development set (OOD-100)
and a disjoint, balanced 1,000-example test set (OOD-1000), each covering signal peptide,
disorder, domain/motif, EC, and GO/InterPro. We retain the benchmark name OOD (out of
distribution) for these held-out sets; the separation from training data is defined below.
The name does not imply that a protein or its annotation is absent from the tools' databases.
The \emph{PFUA-compatible} protocol
uses the Mol-Instructions task categories, prompt, interaction budget, and ROUGE recall scoring
used in prior work~\citep{fan2026pfua,fang2024molinstructions}. We compare scores only within
the same protocol.

\paragraph{Data separation and versions.}
OOD-100 and OOD-1000 share no IDs or sequences. Checkpoints, rewards, and methods are developed
with OOD-100; OOD-1000 is reserved for final reporting. References retain source
Mol-Instructions annotations or deterministic SignalP\,5.0 renderings~\citep{signalp5} and
DisProt annotations. Paper-reported signal-peptide runs execute SignalP\,6.0~\citep{signalp6},
so the label generator and runtime predictor use different versions. This protocol difference may
affect the signal-peptide results, but we have not run a controlled cross-version test. An MMseqs2
audit against the 780-protein
closed-loop training pool finds no detectable homolog at the prespecified 30\% identity /
80\% coverage threshold; over the 2,000-example source superset, 548 sequences fall below that
threshold and 1,452 have no detectable hit. This audit addresses training overlap, not whether a
test protein or homolog is retrievable from a runtime database.

\paragraph{Interaction budgets and baselines.}
The PFUA-compatible protocol permits 12 rounds and 2,048 tokens per response. Closed-loop
evaluation permits $H=4$ interaction steps, 1,024 tokens per response, and 4,096 tokens total.
Closed-loop systems use the same concise protein-function prompt and tool interfaces. We compare
the base and SFT models, the full model with and without tools, four return/normalization
variants, and several frontier API models. The frontier models provide a performance comparison;
the training analysis relies on the matched 8B variants. The local Qwen3 evaluations use
temperature 1.0, top-$p$ 1.0, top-$k$ 50, and seed 42; the API rows in the closed-loop
comparison use temperature 0. The shared protocol therefore fixes the benchmark, prompt, tool
surface, interaction budget, and scorer, but does not impose one decoding configuration across
local and hosted systems.

\subsection{Metrics and Verification}
The primary outcome is annotation-coverage success over all episodes:
$C_{\mathrm{eval}}=R_{\mathrm{tok}}(\operatorname{extract}(y),y^\star)$ computes reference-token
recall over extracted \texttt{<answer>} content, and an episode succeeds when
$C_{\mathrm{eval}}>0.5$. Unanswered and unparsable episodes fail and remain in the denominator.
The scorer follows PFUA's recall construction~\citep{fan2026pfua} and is applied identically to
all systems. At thresholds 0.3--0.7, ProtAgent retains its aggregate advantage over
DeepSeek-V3.2-Thinking, with paired confidence intervals excluding zero.
This metric measures overlap with a reference annotation. It does not directly
measure biological correctness, entailment by tool observations, or calibration.

Answer rate, same-rollout grounded-citation rate, structured-identifier density, and tool-routing
directness are complementary behavioral diagnostics. A high value on any one diagnostic does
not establish correctness or overall grounding. Core comparisons use paired bootstrap intervals
or variation across training seeds as appropriate.

At evaluation time, model parameters are frozen, each benchmark item produces one trajectory,
and no reward is applied. The evaluation code supports four checks:
\emph{tool availability} (required native tools and data are present, with no silent heuristic
fallback), \emph{complete coverage} (one unique rollout per item), \emph{all-sample scoring}
(no failed episode is dropped), and \emph{traceability} (the manifest records prompt, model,
tool/database versions, answer, decoding seed, and code revision).

\section{ProtAgent Post-Training}
\label{sec:method}

\subsection{Data Construction and Supervised Fine-Tuning}
For the closed-loop model, all OOD-100 and OOD-1000 IDs are excluded from SFT and RL.
Demonstrations combine curated action/evidence examples with multi-turn rollouts produced by
DeepSeek-V3.2 using the same bioinformatics tools as the training environment. Thus, the
supervised trajectories contain real tool outputs rather than free-form descriptions. Each
trajectory must have valid tool-call syntax, a substantive answer, and task-critical tool
coverage. The active
SFT corpus contains 987 training and 64 validation trajectories mixing single action examples,
evidence-to-answer examples, and full multi-turn traces (Appendix~\ref{app:training-config}).
During SFT, system and user messages and tool observations are context; only assistant actions
and answers are prediction targets. SFT teaches the interaction interface---tool-call syntax,
task-to-tool routing, and answer format---but the model must still collect evidence at runtime.

For the PFUA-compatible model (Sec.~\ref{sec:pfua}), a separate corpus is
constructed from reviewed Swiss-Prot records~\citep{uniprot} covering the four PFUA task categories.
All Mol-Instructions-derived records, test accessions, and exact test sequences are excluded.
DeepSeek-V3.2-Thinking produces validated tool-use trajectories on 3,900 training
proteins for this model. Because it also appears as a closed-loop baseline
(Sec.~\ref{sec:closed-loop}),
we include three additional frontier systems that provide no training data. SFT teaches the
interface, while RL trains evidence collection, stopping, and answer synthesis on-policy
(Appendix~\ref{app:supplementary}). The two
protocols use separate corpora, rewards, and checkpoints.

\subsection{On-Policy RL in the Protein Environment}
RL as a whole starts from the SFT model and trains it to gather enough evidence, stop at the
right time, and write an answer from the observed results. The paper-facing checkpoint continues
from an existing RL-warm checkpoint through three 50-update warm-restart stages. For each
training problem, the policy samples a group of four trajectories that alternate between a tool
call and the resulting structured observation, following the ReAct interaction
pattern~\citep{react}. The main recipe processes eight prompt instances per update, uses at most
four interaction steps per episode, and uses a constant learning rate within each stage. The
environment rewards useful actions, answer quality, efficient stopping, and optional agreement
with tool outputs. A completed-tool guard prevents format-only rewards when no tool has run.
Training and evaluation use the configured bioinformatics programs and databases; runs with
placeholder outputs are excluded.

The closed-loop RL phase uses three warm-restart stages. These are parts of the RL schedule,
not a third post-training phase. Their learning rates are \(5\times10^{-7}\),
\(5\times10^{-8}\), and \(5\times10^{-8}\), respectively. Each stage loads the previous
checkpoint weights but starts with a fresh optimizer and a new reference anchor. The ablation arms use
the same schedule and starting checkpoint, training questions, sampling schedule, seeds, environment,
reward, budget, and evaluation set. Each arm generates its own trajectories because training is
on-policy. The arms differ in return granularity and return normalization.

\subsection{Composite Reward}
The terminal answer reward is a weighted sum of training-time annotation coverage
$A_{\mathrm{train}}=\min\{1,R_{\mathrm{tok}}(y,y^\star)+0.15z(y)\}$, required-tool coverage $C_{\mathrm{tool}}$,
bounded reasoning markers $R$, efficiency $E$, and structured-answer compatibility $P$, scaled by
a length factor $\ell(y)=\min(1,|y|/200)$:
\begin{equation}
\begin{split}
B &= 2A_{\mathrm{train}} + C_{\mathrm{tool}} + 0.5R + 0.3E + 0.5P,\\
r_{\mathrm{ans}} &= \ell(y)\left[B - 0.45\,\mathbf{1}[m=0] + F\right]\\
&\quad + 0.5J(y,y^\star) + 0.3\,\mathbf{1}[|y|\geq 50 \wedge m\geq 1],
\end{split}
\end{equation}
where $z(y)$ indicates a complete \texttt{<answer>}...\texttt{</answer>} block, $m$ is the number of
completed tools, $J$ is fixed character-Jaccard shaping, and $F$
collects the optional trajectory-local fidelity terms, which are zero in the reward-family
control. The $0.45\,\mathbf{1}[m=0]$ term is the completed-tool guard: it algebraically removes the
schema-only increments that a correctly tagged answer would otherwise earn when no tool has
completed. Intermediate tool calls instead receive action rewards: $+0.25$ for the first
successful use of a new tool, $-0.30$ for a repeated action, $-0.10$ for an irrelevant tool,
$-0.20$ for a malformed call, and $-1.50$ for truncation. Training reward is
used only for optimization diagnostics; all reported comparisons evaluate frozen model
parameters without applying reward at inference time. Appendix~\ref{app:tgsvr} gives the fidelity terms, the full
coefficient table, and a reward-family sensitivity analysis. The coefficient table documents the
implemented objective; it is not itself an ablation.

\subsection{Batched Step-Level REINFORCE}
We use standard critic-free, on-policy REINFORCE~\citep{williams1992reinforce}. For each prompt,
Batched Step-Level
REINFORCE collects $K=4$ trajectories and applies Monte Carlo returns~\citep{sutton2018reinforcement} at
each interaction step, rather than assigning one terminal score to every action. Let $r_{t,k}$
be the reward at step $t$ of trajectory $k$. It is the action reward for an intermediate tool
call and the composite answer reward for the final response. We compute returns backward:
\begin{equation}
G_{t,k} = r_{t,k} + \gamma G_{t+1,k}, \qquad G_{H+1,k}=0,
\end{equation}
with $\gamma=1$ in the main configuration. The main recipe uses identity return normalization.
All comparison arms then whiten and clip the resulting advantages before one policy-gradient
update. Each response receives
its step return, and later rewards propagate only to earlier actions in the same trajectory.
With a fixed reference
policy, the actor objective is
\begin{equation}
\begin{aligned}
\mathcal{L}_{\mathrm{RL}}
&= -\mathbb{E}_{t,u}\!\left[\widehat{A}_{t,u}
\log \pi_\theta(y_{t,u}\mid h_{t,u})\right] \\
&\quad + \beta D_{\mathrm{KL}}\!\left(\pi_\theta\,\|\,\pi_{\mathrm{ref}}\right),
\end{aligned}
\end{equation}
where $u$ indexes tokens within the response at step $t$; Appendix~\ref{app:training-config}
lists the remaining coefficients. The matched episode-level baseline keeps the same setup but
assigns one trajectory-level return to all actions in each independently sampled rollout. Our
step return differs from the state-level group term in GiGPO~\citep{gigpo} and from the
group-relative turn rewards in MT-GRPO~\citep{wei2025mtgrpo}: it follows rewards within one
trajectory. We vary return granularity and return normalization separately
(Table~\ref{tab:optimization-ablation}). We use \emph{Step-Level GRPO} for step-level
Monte Carlo returns standardized by the within-prompt group mean and standard deviation.
This label changes return normalization only: the loss remains the same vanilla policy-gradient
objective, not the framework's episode-level \texttt{adv\_estimator: grpo}. Episode-GRPO applies
the same group standardization to episode-level returns~\citep{deepseekmath}.

\section{Experiments}

\subsection{Closed-Loop Tool-Use Evaluation}
\label{sec:closed-loop}
We first report the main closed-loop experiment. It uses a fixed local checkpoint, pinned tool
versions, complete benchmark coverage, and five annotation tasks. This experiment tests tool
selection, use of tool outputs, and stopping under the protocol in Sec.~\ref{sec:protocol}.

\subsubsection{Compact versus frontier-scale agents}
Table~\ref{tab:headline} reports the main closed-loop comparison.
\begin{table*}[t]
\centering
{\small
\begin{tabularx}{\textwidth}{@{}l*{6}{>{\centering\arraybackslash}X}@{}}
\toprule
\textbf{Method} & \textbf{Enzyme} & \textbf{Domain} & \textbf{GO/IPR} & \textbf{Signal} & \textbf{Disorder} & \textbf{Aggregate} \\
\midrule
DeepSeek-V3.2-Thinking 671B & 1.00 & 25.50 & 1.50 & \textbf{99.00} & 81.00 & 41.60 \\
DeepSeek-V4-Flash & 7.00 & 29.00 & 42.50 & \textbf{99.00} & \textbf{94.50} & 54.40 \\
DeepSeek-V4-Pro & 14.50 & 30.00 & 27.00 & 98.50 & 92.00 & 52.40 \\
GPT-5.5 & 53.50 & \textbf{35.00} & 43.00 & \textbf{99.00} & 73.50 & 60.80 \\
\rowcolor{rowOurs}ProtAgent-8B (closed-loop) & \textbf{58.00} & \textbf{35.00} & \textbf{54.50} & 73.50 & 84.00 & \textbf{61.00} \\
\bottomrule
\end{tabularx}}
\caption{All-episode success (\%) on the fixed OOD-1000 protocol (1{,}000 sequences; 200 per task
category). Boldface marks the best value in each column. All systems use the same concise
protein-function prompt and the same bioinformatics tools. Unanswered episodes count as
failures. Each run uses a fixed decoding configuration. The rates describe the observed outputs
on this test set and do not guarantee identical API outputs across repeated runs.}
\label{tab:headline}
\end{table*}

Against DeepSeek-V3.2-Thinking, the paired aggregate difference is $+19.40$ percentage points
(95\% bootstrap CI [$+15.90$, $+23.00$]; 10{,}000 paired resamples).
ProtAgent-8B scores higher than both evaluated DeepSeek-V4 variants and close to
GPT-5.5 (61.00\% vs.\ 60.80\%); this small difference does not establish superiority or
statistical equivalence. Three frontier systems in the table provide no training data.
The gain over DeepSeek-V3.2-Thinking is concentrated in enzyme, domain, and GO/IPR;
the baseline remains better on signal peptide. Overall, these results show that the compact
local model is competitive with much larger hosted systems under the same closed-loop protocol,
while revealing clear task-level differences.

\subsection{Training-Stage and Tool Controls}
Table~\ref{tab:stage-controls} compares training stages and evaluates the full checkpoint with
and without tool access, by OOD-1000 task category.

\begin{table}[t]
\centering
{\small
\begin{tabular*}{\columnwidth}{@{\extracolsep{\fill}}lcccc@{}}
\toprule
Category & Base & SFT & Full & No tools \\
\midrule
Enzyme & 3.00 & 23.50 & \textbf{58.00} & 11.50 \\
Domain & 0.50 & 17.00 & \textbf{35.00} & 1.50 \\
GO/IPR & 7.00 & 17.50 & \textbf{54.50} & 10.50 \\
Signal & 51.00 & \textbf{78.50} & 73.50 & 31.00 \\
Disorder & 9.00 & 64.50 & \textbf{84.00} & 73.00 \\
\midrule
Aggregate & 14.10 & 40.20 & \textbf{61.00} & 25.50 \\
\bottomrule
\end{tabular*}}
\caption{Per-category OOD-1000 success (\%) for training-stage and tool-access controls.
Full = SFT + step-level RL; No tools = the full checkpoint with tools ablated.}
\label{tab:stage-controls}
\end{table}

On the OOD-100 development set, a matched same-tools comparison changes only the checkpoint.
SFT scores 15.00\%, while the full model scores 57.00\%, a paired difference of 42.00 points
(95\% CI [31.00, 53.00]). Both results come from the same validation round and are paired by
sample. SFT and RL mainly improve the database-based tasks; removing the tools lowers performance
on those tasks toward the base model.

\subsection{Credit-Assignment Study}
\paragraph{Experimental controls.}
OOD-1000 annotations are not used during training, and no reward is applied during evaluation.
The credit-assignment arms share the same starting checkpoint, training questions, sampling schedule,
seeds, environment, reward, budget, and evaluation set. Because training is on-policy, each arm
samples its own trajectories. The arms differ only in return granularity and return
normalization.

\paragraph{Answer completion.}
Answer rate is 98.50\% for the trained 8B model and 99.60\% for the tool-powered baseline.
The higher success rate of the trained model therefore does not come from a higher answer rate.

\subsubsection{Return granularity and return normalization}
We compare step- and episode-level returns under both identity and within-prompt group
return normalization. The environment, reward, prompt groups, and training budget are fixed.

\begin{figure}[t]
\centering
\includegraphics[width=\columnwidth]{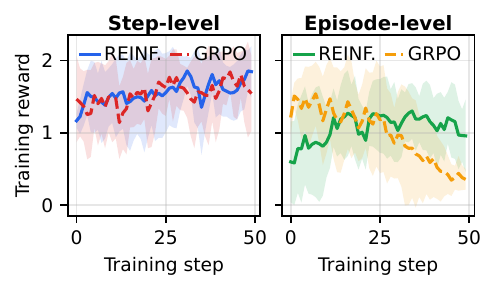}
\caption{Training-reward trajectories grouped by return granularity (three-seed mean,
$\pm1$ standard-deviation band, light exponential smoothing). Solid lines denote REINFORCE and
dashed lines denote GRPO. Reward is an optimization diagnostic rather than a cross-arm quality
metric; Table~\ref{tab:optimization-ablation} reports final-test success.}
\label{fig:training-curves}
\end{figure}

\begin{table}[t]
\centering
{\footnotesize
\setlength{\tabcolsep}{3pt}%
\begin{tabularx}{\columnwidth}{@{}Ylc@{}}
\toprule
Arm & Return norm. & Mean\,$\pm$\,SD \\
\midrule
Batched Step-Level REINFORCE & Identity & $61.40\pm2.10$ \\
Step-Level GRPO & Group-rel. & $58.20\pm1.70$ \\
Batched Episode-Level REINFORCE & Identity & $42.70\pm1.60$ \\
Episode-GRPO & Group-rel. & $36.50\pm1.70$ \\
\bottomrule
\end{tabularx}}
\caption{OOD-1000 all-episode success (\%), mean $\pm$ sample SD over seeds 42/43/44.
Arms share the same starting checkpoint, training questions, sampling schedule, seeds, environment, reward,
budget, and evaluation set. Each arm produces its own on-policy trajectories. Appendix
\ref{app:tgsvr} gives per-seed values and the separate reward-family sensitivity analysis.}
\label{tab:optimization-ablation}
\end{table}

In Table~\ref{tab:optimization-ablation}, step-level returns score $+18.70$ points above
episode-level returns under identity return normalization and $+21.70$ under group-relative
return normalization. These large gains have the same direction across all three seeds, supporting
interaction-level credit assignment as the main design choice in this training setup.
At episode level, identity return normalization scores $+6.20$ points above group-relative
return normalization, with the
same ordering across seeds. The reward-family on/off runs in the supplement were not
produced as one concurrent matched experiment. We report them as a sensitivity check and do not
use them to estimate the effect of the optional fidelity reward.
The step-level-return arms show more stable training-reward trajectories, whereas the
episode-level arms decline late in training, most clearly under Episode-GRPO. These curves
describe training reward, not held-out answer quality.

\subsection{Task and Failure Analysis}
Most of the aggregate gain comes from database-based tasks. Signal-peptide prediction remains a
weak point. Its labels were rendered with SignalP\,5.0~\citep{signalp5}, while the runtime tool
uses SignalP\,6.0~\citep{signalp6}. This version difference may matter, but we do not have a matched
cross-version experiment.
The aggregate grounded-citation rate is lower for ProtAgent (76.10\% vs.\ 81.50\%), mainly
because of signal and disorder tasks. On database tasks, the rates are similar for enzyme
(90.00\% vs.\ 92.00\%) and higher for ProtAgent on domain (93.00\% vs.\ 79.00\%) and GO/IPR
(93.00\% vs.\ 81.00\%). Across these three database tasks, ProtAgent also produces more
structured identifiers per answer (2.5 vs.\ 0.5) and chooses a relevant tool on the first call
more often (98.50\% vs.\ 2.00\%). Both models reach a relevant tool by some point in almost every
trajectory (99.50\% vs.\ 100.00\%). Routing and identifier counts describe model behavior, not
biological correctness.

Appendix~\ref{app:supplementary} reports the full category and diagnostic panels and defines a failure taxonomy
covering tool selection, tool failure, premature stopping, evidence omission, unsupported claims,
and formatting or truncation. This taxonomy defines categories for error analysis; it is not
a completed annotation study, and we report no category counts.

\subsection{PFUA-Compatible External Evaluation}
\label{sec:pfua}
Finally, we evaluate the separately trained PFUA-compatible model on the Mol-Instructions
protein-function benchmark~\citep{fan2026pfua,fang2024molinstructions}. Table~\ref{tab:pfua-compact}
shows the strongest tool-powered result published by PFUA, the strongest frontier result from
our reproduction, and the Qwen3-8B training stages. Published results and our reproductions are
labeled separately because they were run at different times and with different model endpoints.
All rows marked ``evaluated by us'' use the same released PFUA subset, prompt, tool interfaces,
interaction budget, output processing, and official scorer.
Appendix Table~\ref{tab:pfua-complete} gives the full comparison, and
Appendix~\ref{app:reproduction-variance} gives the reproduction details. The PFUA-compatible model is
not the closed-loop checkpoint, so its 60.15/41.62 score is reported only for comparison with
PFUA.

\begin{table}[t]
\centering
{\small
\setlength{\tabcolsep}{4pt}%
\begin{tabular}{@{}lcc@{}}
\toprule
System and provenance & R-1 & R-L \\
\midrule
Qwen3-Max (PFUA-reported) & 64.33 & 46.16 \\
GPT-5.5 (evaluated by us) & \textbf{61.41} & \textbf{44.66} \\
\midrule
Qwen3-8B base (evaluated by us) & 53.62 & 35.33 \\
Qwen3-8B SFT (evaluated by us) & 54.47 & 37.68 \\
\rowcolor{rowOurs}ProtAgent-8B PFUA (evaluated by us) & 60.15 & 41.62 \\
\bottomrule
\end{tabular}}
\caption{Compact PFUA-compatible comparison: average ROUGE-1/ROUGE-L recall (\%). The PFUA
published row is prior-work context and is not ranked with our runs. Bold marks the best result
from our evaluation. The PFUA-compatible ProtAgent uses a different dataset, reward, and
checkpoint from the closed-loop model.}
\label{tab:pfua-compact}
\end{table}

In our PFUA-compatible evaluation, the model improves from 53.62/35.33 for the prompted base and
54.47/37.68 after SFT to 60.15/41.62 after RL, exceeding five of six evaluated frontier APIs in
the full panel. DeepSeek and Kimi reproduce within 0.7 and 0.3 average ROUGE-1 points,
respectively. The current Qwen3-Max endpoint is 7.5 points below the published preview result,
so comparisons within our evaluation use only the reproduced rows.

\section{Conclusion}
Under a fixed tool-use protocol, a post-trained local 8B model outperforms the evaluated
DeepSeek-V3.2-Thinking 671B baseline and obtains a score close to the strongest evaluated
frontier system on aggregate annotation-coverage success. The 19.40-point gain over the 671B
baseline and the 61.00\% aggregate score show that compact local agents can be competitive with
frontier-scale systems when they are post-trained in the target tool environment. In the
three-seed comparison, step-level returns provide large, consistently directed gains over
episode-level returns under both normalization methods.
The full model scores higher than the base and SFT models, and its performance drops when tools
are removed. On the database-based tasks, it also selects relevant tools earlier and produces
more structured identifiers. In this training setup, interaction-level returns and
environment-specific post-training improve a compact protein tool agent. The results do not
establish biological correctness or stronger overall evidence grounding.

\section*{Limitations}
\paragraph{Scope.}
Our conclusions are limited to protein annotation with the evaluated Qwen3-8B model, tools,
benchmarks, and baselines. The frontier-model comparison uses one shared prompt, tool setup,
interaction budget, scorer, and a fixed decoding configuration for each run. It does not isolate
model architecture from other system-level differences, including decoding choices, and is not
a general ranking of model architectures.
The credit-assignment comparison is specific to the evaluated training setup, not a general
comparison of RL algorithms.

\paragraph{Evidence quality and robustness.}
Computational tool evidence is not equivalent to experimental validation, and its reliability
varies across tasks and protein families. Our homology audit checks train--test overlap at a
fixed threshold, but it cannot rule out remote relationships or determine whether each test
annotation is available in the runtime databases. The main scorer measures reference-annotation
overlap, while the citation metric measures support within the observed trajectory.
Neither substitutes for expert biological assessment. Without a deterministic tool-output
renderer or an oracle-observation baseline, we cannot separate gains in evidence acquisition
from gains in answer construction. SignalP\,5.0 labels~\citep{signalp5} and SignalP\,6.0 runtime
predictions~\citep{signalp6} also differ in version; without a matched test, we cannot attribute
the signal-peptide deficit to this difference. Finally, the non-concurrent fidelity-reward
comparison does not isolate its effect, and the contribution of each reward component remains
untested.

\section*{Acknowledgments}
Generative AI tools were used during manuscript revision to support language editing,
organization, and literature search. The authors verified the resulting text, citations,
technical claims, and reported results and take full responsibility for the submission.

\bibliography{references}

\clearpage
\appendix
\renewcommand{\thetable}{\thesection\arabic{table}}
\renewcommand{\thefigure}{\thesection\arabic{figure}}
\makeatletter\@addtoreset{table}{section}\@addtoreset{figure}{section}\makeatother
\section{Reward Definition and Trajectory-Fidelity Sensitivity}
\label{app:tgsvr}

\paragraph{Reward definition.}
Tool-Grounded Self-Verifiable Reward (\tgsvr{}) is an optional fidelity family inside the
composite training objective; it is not the entire reward and is not required by the evaluation
metric. For a trajectory $\tau$
with final answer $y$ and $m$ completed tools, the selected recipe assigns
\begin{equation}
\begin{split}
B(y,\tau) &= 2A_{\mathrm{train}} + C_{\mathrm{tool}}\\
&\quad + 0.5R + 0.3E + 0.5P,\\
r_{\mathrm{ans}} &= \ell(y)\left[B(y,\tau)-0.45\mathbf{1}[m=0]\right.\\
&\quad\left.+0.2F_{\mathrm{dis}}+1.0F_{\mathrm{cat}}\right]\\
&\quad +0.5J(y,y^\star)\\
&\quad +0.3\mathbf{1}[|y|\geq50\wedge m\geq1],
\end{split}
\label{eq:composite-reward}
\end{equation}
where $A_{\mathrm{train}}$ is the training-time answer-coverage term (including the formatting
bonus), and $C_{\mathrm{tool}},R,E,P$ denote required-tool coverage, bounded reasoning markers, efficiency, and
structured-answer compatibility; $J$ is fixed character-Jaccard shaping and
$\ell(y)=\min(1,|y|/200)$. $F_{\mathrm{dis}}$ compares predicted disorder ranges with the
same-rollout disorder observation, while $F_{\mathrm{cat}}$ compares EC, SignalP, Pfam, or
GO/IPR claims with their corresponding same-rollout fields. The no-TGSVR arms set only these two
coefficients to zero. Table~\ref{tab:reward-weights} lists the full coefficient set.

\paragraph{Three-seed results.}
The On rows are the four-arm, three-seed comparison reported in the main text. The Off rows are
from earlier experiments. The available run records do not show that every On and Off run used
the same stabilized training configuration during the same experimental period. We therefore
use the on/off comparison only as a sensitivity check; it is neither a matched factorial
experiment nor an estimate of the effect of TGSVR. The main result uses the On rows to compare
return granularity and return normalization. The citation,
identifier-density, and routing measures in Appendix~\ref{app:supplementary} describe model
behavior rather than biological validity.

\begin{table*}[t]
\centering
{\small
\begin{tabular}{@{}lllccccc@{}}
\toprule
\multirow{2}{*}{Arm} & \multirow{2}{*}{Return} & \multirow{2}{*}{Return norm.} & \multirow{2}{*}{TGSVR} & \multicolumn{3}{c}{Seed} & \multirow{2}{*}{Mean\,$\pm$\,SD} \\
\cmidrule{5-7}
& & & & 42 & 43 & 44 \\
\midrule
\multirow{2}{*}{Batched Step-Level REINFORCE} & \multirow{2}{*}{Step} & \multirow{2}{*}{Identity} & On & 63.40 & 59.20 & 61.60 & $61.40\pm2.10$ \\
& & & Off & 62.10 & 58.90 & 61.10 & $60.70\pm1.60$ \\
\addlinespace[2pt]
\multirow{2}{*}{Step-Level GRPO} & \multirow{2}{*}{Step} & \multirow{2}{*}{Group-rel.} & On & 59.80 & 56.50 & 58.30 & $58.20\pm1.70$ \\
& & & Off & 59.20 & 56.80 & 57.80 & $57.90\pm1.20$ \\
\addlinespace[2pt]
\multirow{2}{*}{Batched Episode-Level REINFORCE} & \multirow{2}{*}{Episode} & \multirow{2}{*}{Identity} & On & 44.20 & 41.00 & 42.90 & $42.70\pm1.60$ \\
& & & Off & 43.50 & 41.30 & 42.50 & $42.40\pm1.10$ \\
\addlinespace[2pt]
\multirow{2}{*}{Episode-GRPO} & \multirow{2}{*}{Episode} & \multirow{2}{*}{Group-rel.} & On & 38.20 & 34.80 & 36.50 & $36.50\pm1.70$ \\
& & & Off & 37.50 & 35.20 & 36.10 & $36.30\pm1.20$ \\
\bottomrule
\end{tabular}}
\caption{Reported three-seed return-granularity and return-normalization results with TGSVR
sensitivity rows. Entries are
all-episode OOD-1000 success (\%); the final column reports the across-seed mean $\pm$ sample
standard deviation. On rows are the main-text stabilized retrains. Off rows retain the original
reported values from earlier experiments, but they were not produced as part of the same matched
experiment. The differences between On and Off therefore do not isolate the effect of TGSVR.}
\label{tab:tgsvr-factorial}\end{table*}

\begin{table*}[t]
\centering{\small
\setlength{\tabcolsep}{4pt}%
\begin{tabularx}{\textwidth}{p{0.29\textwidth}cp{0.23\textwidth}Y}
\toprule
Component & Weight & Applied at & Role \\
\midrule
Answer coverage & $2.0$ & Terminal answer & Reference-annotation recovery shaping \\
Required-tool coverage & $1.0$ & Terminal answer & Task-relevant tool completion \\
Reasoning / efficiency / structure & $0.5/0.3/0.5$ & Terminal answer & Bounded behavior and schema terms \\
Similarity $J$ & $0.5$ & Terminal answer & Training-only character-Jaccard shaping \\
Completed-tool guard & $-0.45$ if $m=0$ & Terminal answer & Removes schema-only reward without a completed tool \\
Trajectory-local fidelity & $0.2/1.0$ & Terminal answer & Disorder / categorical agreement with same-rollout observations \\
First new correct tool & $+0.25$ & Intermediate action & Successful first use of a relevant tool \\
Repeated / irrelevant / malformed & $-0.30/-0.10/-0.20$ & Intermediate action & Action-level penalties \\
Truncation / valid-answer bonus & $-1.50/+0.30$ & Boundary / terminal & Truncation penalty and tool-gated answer completion \\
\bottomrule
\end{tabularx}}
\caption{Composite reward components and attachment points. This table documents the implemented
objective; it is not a component ablation. The completed-tool correction is algebraic:
$0.45=2(0.15)+0.5(0.30)$ removes schema-only increments when no tool completes.}
\label{tab:reward-weights}
\end{table*}

\paragraph{Comparison scope.}
The main-text On arms share a common starting checkpoint, training questions, sampling schedule, seeds,
environment, reward, budget, and evaluation set. Each arm generates its own on-policy
trajectories. They differ in return granularity and return normalization. A causal test of individual
reward components would require matched runs that remove one component at a time. The retained
Off rows do not meet that requirement.

\section{Supplementary Analyses}
\label{app:supplementary}

\paragraph{Training-stage and tool ablation.}
Table~\ref{tab:tool-training-controls} separates two factors: the training stage that produced
the policy and the availability of bioinformatics tools at evaluation time. Aggregate all-episode success
rises from 14.10 for the Qwen3-8B base through 40.20 after supervised fine-tuning to 61.00 for the
full method, and the increase is concentrated in the three database-evidence categories rather than
spread evenly across the five. Removing the actual tools from the full-method checkpoint returns
the aggregate to 25.50---below its supervised-only stage. This comparison measures the effect of
tool availability for the full checkpoint; it does not separate evidence acquisition from answer
construction.

\begin{table*}[t]
\centering
{\small
\begin{tabularx}{\textwidth}{@{}l*{6}{>{\centering\arraybackslash}X}@{}}
\toprule
\textbf{System} & \textbf{Enzyme} & \textbf{Domain} & \textbf{GO/IPR} & \textbf{Signal} & \textbf{Disorder} & \textbf{Aggregate} \\
\midrule
Qwen3-8B (base) & 3.00 & 0.50 & 7.00 & 51.00 & 9.00 & 14.10 \\
ProtAgent-8B (SFT) & 23.50 & 17.00 & 17.50 & 78.50 & 64.50 & 40.20 \\
\rowcolor{rowOurs}ProtAgent-8B (full method) & 58.00 & 35.00 & 54.50 & 73.50 & 84.00 & 61.00 \\
ProtAgent-8B (tools ablated) & 11.50 & 1.50 & 10.50 & 31.00 & 73.00 & 25.50 \\
\bottomrule
\end{tabularx}}
\caption{Training-stage and tool ablation (all-episode success, \%) on the fixed OOD-1000
protocol, reported per task
category. Rows trace the pipeline from the Qwen3-8B
base through supervised fine-tuning (SFT), the full method (SFT with step-level RL), and a
tools-ablated control. All rows except the tools-ablated control use the same bioinformatics
tools. The tools-ablated control intentionally receives no tool observations.
The matched OOD-100 same-tools control (SFT $\to$ full method) is reported in the main text.}
\label{tab:tool-training-controls}
\end{table*}

\paragraph{Behavioral diagnostics.}
Table~\ref{tab:evidence-panel} compares the two systems on tool routing, answer rate, citations,
and structured identifiers. Both systems eventually reach a relevant tool on almost every
example. The main differences are earlier routing and identifier use: ProtAgent selects a
relevant tool more often on the first call and reports 1.7 database accessions per answer,
compared with 0.5 for the baseline. These measures describe behavior, not whether every cited
identifier supports every claim. ProtAgent's aggregate grounded-citation rate is also lower than
the baseline's (76.10\% versus 81.50\%).

\begin{table*}[!t]
\centering
{\small
\setlength{\tabcolsep}{4pt}
\begin{tabularx}{\textwidth}{@{}l*{6}{>{\centering\arraybackslash}X}@{}}
\toprule
\textbf{Diagnostic / system} & \textbf{Enzyme} & \textbf{Domain} & \textbf{GO/IPR} & \textbf{Signal} & \textbf{Disorder} & \textbf{Aggregate} \\
\midrule
\multicolumn{7}{@{}l}{\textit{All-episode success} --- primary outcome}\\
\rowcolor{rowOurs}\quad ProtAgent-8B & 58.00 & 35.00 & 54.50 & 73.50 & 84.00 & 61.00 \\
\quad DeepSeek-V3.2-Thinking & 1.00 & 25.50 & 1.50 & 99.00 & 81.00 & 41.60 \\
\midrule
\multicolumn{7}{@{}l}{\textit{Answer rate} --- stopping behavior}\\
\rowcolor{rowOurs}\quad ProtAgent-8B & 99.00 & 99.00 & 99.00 & 99.00 & 96.50 & 98.50 \\
\quad DeepSeek-V3.2-Thinking & 100.00 & 100.00 & 100.00 & 99.00 & 99.00 & 99.60 \\
\midrule
\multicolumn{7}{@{}l}{\textit{Grounded-citation rate} --- traceability of claims}\\
\rowcolor{rowOurs}\quad ProtAgent-8B & 90.00 & 93.00 & 93.00 & 50.00 & 54.50 & 76.10 \\
\quad DeepSeek-V3.2-Thinking & 92.00 & 79.00 & 81.00 & 56.00 & 99.50 & 81.50 \\
\midrule
\multicolumn{7}{@{}l}{\textit{Structured identifiers per answer} --- evidence density (count, not a rate)}\\
\rowcolor{rowOurs}\quad ProtAgent-8B & 2.2 & 3.7 & 1.6 & 0.5 & 0.6 & 1.7 \\
\quad DeepSeek-V3.2-Thinking & 0.6 & 0.5 & 0.4 & 0.3 & 0.7 & 0.5 \\
\midrule
\multicolumn{7}{@{}l}{\textit{Skill-relevant tool routing, first call} --- routing efficiency}\\
\rowcolor{rowOurs}\quad ProtAgent-8B & 100.00 & 99.00 & 96.50 & 35.00 & 100.00 & 98.50$^{\dagger}$ \\
\quad DeepSeek-V3.2-Thinking & 0.00 & 1.00 & 5.00 & 28.00 & 0.00 & 2.00$^{\dagger}$ \\
\midrule
\multicolumn{7}{@{}l}{\textit{Skill-relevant tool routing, any step} --- routing capability (ceiling control)}\\
\rowcolor{rowOurs}\quad ProtAgent-8B & 100.00 & 100.00 & 98.50 & 98.00 & 100.00 & 99.50$^{\dagger}$ \\
\quad DeepSeek-V3.2-Thinking & 100.00 & 100.00 & 100.00 & 100.00 & 100.00 & 100.00$^{\dagger}$ \\
\bottomrule
\end{tabularx}}
\caption{Behavioral diagnostics on OOD-1000. All values are percentages except identifiers per
answer. Unless marked, aggregates are macro means over all five tasks; $^{\dagger}$ indicates an
aggregate over the three database-based tasks only. Citation rates and identifier counts use
answered episodes. Identifiers are database accessions (EC, Pfam, GO, or UniProt), not residue
ranges. The same relevant-tool sets are used for both systems. The any-step routing measure is a
ceiling check: reaching a relevant tool does not guarantee that the returned evidence is useful
or that the final answer uses it correctly.}
\label{tab:evidence-panel}
\end{table*}

\paragraph{Failure taxonomy.}
Table~\ref{tab:failure-taxonomy} defines categories for analyzing unsuccessful trajectories.
The primary label denotes the first failure that prevents a supported, task-compliant answer.
Later problems receive secondary labels. We have not completed the annotation and adjudication
needed to estimate category frequencies, so no counts are reported.

\begin{table*}[!t]
\centering
{\small
\setlength{\tabcolsep}{5pt}%
\begin{tabularx}{\textwidth}{@{}p{0.18\textwidth}p{0.33\textwidth}Y@{}}
\toprule
Primary category & Operational definition & Audit evidence \\
\midrule
Tool selection & No task-relevant tool is called before termination, or the selected tool cannot
produce evidence for the requested annotation. & Ordered tool calls, task-to-tool eligibility map,
and termination state. \\
Tool failure & A relevant call is attempted but execution, parsing, dependency, or database access
fails. & Tool status, stderr/error field, observation schema, and native-backend manifest. \\
Premature stopping & The policy emits a terminal answer while an indicated evidence gap remains,
or before a necessary completed observation is available. & Final action position, completed-call
set, and task-required evidence fields. \\
Evidence omission & Relevant evidence is present in an actual observation but the final answer
omits the annotation needed by the reference-coverage scorer. & Same-trajectory observations,
extracted answer, and reference annotation. \\
Unsupported claim & The final answer states an identifier, range, or function that is absent from
or contradicted by the trajectory's observations. & Claim-to-observation alignment with exact
identifier and span matching, followed by manual adjudication. \\
Formatting / truncation & Content cannot be scored because the answer wrapper or required schema
is malformed, or generation ends at the response/token limit. & Parser outcome, finish reason,
response length, and raw terminal output. \\
\bottomrule
\end{tabularx}}
\caption{Failure categories and the evidence needed to assign them. The table defines an
annotation scheme, not the results of a completed error-labeling study.}
\label{tab:failure-taxonomy}
\end{table*}

\paragraph{Credit-assignment variants.}
Table~\ref{tab:credit-assignment} varies token-level advantage mixing, step discounting, and
step eligibility traces around Batched Step-Level REINFORCE. Each row is a fixed-budget
continuation of one shared step-level checkpoint on OOD-100.

\begin{table}[t]
\centering
{\small
\begin{tabularx}{\columnwidth}{Yc}
\toprule
\textbf{Credit-assignment setting} & \textbf{OOD-100 success (\%)} \\
\midrule
\textit{Continuation control (default, $\lambda=0.95$)} & 51.00 \\
\midrule
\multicolumn{2}{@{}l}{\textit{Token-level}} \\
\quad $\lambda=0.3$ & 48.00 \\
\quad $\lambda=0.5$ & 41.00 \\
\quad $\lambda=0.7$ & 51.00 \\
\quad $\lambda=0.9$ & 52.00 \\
\midrule
\multicolumn{2}{@{}l}{\textit{Step discount}} \\
\quad $\gamma=0.90$ & 46.00 \\
\quad $\gamma=0.95$ & 50.00 \\
\midrule
\multicolumn{2}{@{}l}{\textit{Step eligibility trace}} \\
\quad $\lambda=0.5$ & 52.00 \\
\quad $\lambda=0.7$ & 55.00 \\
\bottomrule
\end{tabularx}}
\caption{Credit-assignment ablations around Batched Step-Level REINFORCE on OOD-100 under a
shared tool environment. All rows use ProtAgent-8B and continue from one shared step-level
checkpoint under a single evaluation protocol. The control is the continuation baseline of
this sweep, not the separately evaluated main-text model. Rows are comparable within the table
but not to the main-text OOD-100 value.
Success is computed over all 100 examples, with unanswered episodes counted as failures.
No setting with paired sample-level records differs significantly from the continuation control.}
\label{tab:credit-assignment}
\end{table}

Across the six settings with paired sample-level records, Cochran's
$Q=1.771$ ($p=0.880$), and none of the 15 pairwise exact McNemar tests is significant
($p>0.35$). These tests do not detect a difference among the evaluated settings; they do not
establish equivalence or support a ranking.

\paragraph{Extended algorithm comparison.}
Table~\ref{tab:extended-algo-ablation} combines the main comparison with an earlier single-seed
study of three additional estimators. These are a PPO-clipped objective~\citep{ppo} on the
step-level return, a GiGPO-style dual-level advantage, and episode-level agentic REINFORCE.
The two studies use different training configurations and evaluation sets. We retain the earlier
results as context, not as a controlled ranking.

\begin{table*}[!t]
\centering
{\small
\setlength{\tabcolsep}{5pt}
\begin{tabular}{@{}lllcc@{}}
\toprule
Arm & Return & Return norm. & Held-out reward & All-episode success (\%) \\
\midrule
Batched Step-Level REINFORCE$^{\dagger}$ & Step & Identity & $2.313\pm0.084$ & $61.40\pm2.10$ \\
Step-Level GRPO$^{\dagger}$ & Step & Group-rel. & $2.023\pm0.106$ & $58.20\pm1.70$ \\
Batched Step-Level REINFORCE + PPO clip & Step & Identity & $1.847\pm0.091$ & $47.80\pm1.80$ \\
GiGPO-style dual-level & Step+Ep. & Dual & $1.893\pm0.117$ & $45.60\pm2.60$ \\
Batched Episode-Level REINFORCE$^{\dagger}$ & Episode & Identity & $1.333\pm0.073$ & $42.70\pm1.60$ \\
Episode-Level Agentic REINFORCE & Episode & Identity & $0.902\pm0.058$ & $13.70\pm1.10$ \\
Episode-GRPO$^{\dagger}$ & Episode & Group-rel. & $0.737\pm0.062$ & $36.50\pm1.70$ \\
\bottomrule
\end{tabular}}
\caption{Algorithm results from two experimental settings. For $\dagger$ rows, success is
the main-text OOD-1000 mean $\pm$ sample SD across training seeds 42/43/44. The three unmarked
rows are single-seed OOD-100 evaluations from an earlier training configuration; their
$\pm$ values describe evaluation sampling variability, not variation across training seeds.
All reward entries come from the earlier single-seed experiments and are not co-sampled with
the $\dagger$ success entries. Reward is a diagnostic under the composite objective
(Appendix~\ref{app:tgsvr}), not a shared scale for ranking these experiments.}
\label{tab:extended-algo-ablation}
\end{table*}

\paragraph{Threshold sensitivity.}
The headline reports the 0.5 reference-token annotation-coverage operating point. To test whether
the comparison depends on that choice, we re-score
the fixed OOD-1000 outputs at five cutoffs $\tau \in \{0.3, 0.4, 0.5, 0.6, 0.7\}$ and compute
paired bootstrap 95\% intervals (10{,}000 resamples) for the ProtAgent-minus-baseline difference.
Table~\ref{tab:threshold-sweep} shows that ProtAgent-8B exceeds the tool-powered baseline
at every threshold, with all confidence intervals excluding zero. The advantage ranges from
$+6.40$ percentage points at $\tau{=}0.7$ to $+29.30$ at $\tau{=}0.4$. Thus, the advantage over
this baseline is not limited to the 0.5 cutoff within the tested range.

\begin{table}[!b]
\centering
{\footnotesize
\setlength{\tabcolsep}{3pt}%
\begin{tabular}{@{}crrrc@{}}
\toprule
$\tau$
& \multicolumn{1}{c}{\shortstack{ProtAgent-\\8B}}
& \multicolumn{1}{c}{\shortstack{DeepSeek-\\V3.2-\\Thinking}}
& \multicolumn{1}{c}{$\Delta$}
& 95\% CI \\
\midrule
0.30 & 87.20 & 60.20 & +27.00 & [+23.90, +30.20] \\
0.40 & 78.30 & 49.00 & +29.30 & [+25.90, +32.80] \\
0.50 & 61.00 & 41.60 & +19.40 & [+15.90, +23.00] \\
0.60 & 40.90 & 23.70 & +17.20 & [+14.00, +20.40] \\
0.70 & 22.90 & 16.50 & +6.40 & [+3.50, +9.20] \\
\bottomrule
\end{tabular}}
\caption{Aggregate all-episode success (\%) on OOD-1000 at varying reference-token
annotation-coverage cutoffs
$\tau$; $\Delta$ and its interval are in percentage points.
ProtAgent-8B exceeds the DeepSeek-V3.2-Thinking baseline at every threshold; all paired bootstrap
95\% intervals (10{,}000 resamples) exclude zero. The disorder category uses the compaction-fixed
evaluation; the remaining four categories carry over from the primary evaluation.}
\label{tab:threshold-sweep}
\end{table}

\paragraph{Trajectory-source dual role.}
The DeepSeek-V3.2-Thinking model that produces the PFUA-layer supervised trajectories is also one
of the evaluated baselines. We state both roles explicitly rather than treating it as a fully
independent baseline. The closed-loop table also includes DeepSeek-V4-Flash, DeepSeek-V4-Pro, and
GPT-5.5, none of which provides training data. In addition, the SFT trajectories mainly teach the
tool interface, while RL trains evidence collection, stopping, and answer synthesis on-policy.
All four credit-assignment arms start from the same SFT checkpoint, keeping the demonstration
source fixed within that comparison.

\section{PFUA Complete Results and Reproduction Analysis}
\label{app:reproduction-variance}

Table~\ref{tab:pfua-complete} gives the complete tool-powered comparison summarized in the main
paper. It separates results published by PFUA, frontier models evaluated by us, and checkpoints
trained by us. Table~\ref{tab:pfua-smallscale} lists the remaining PFUA results.

\begin{table*}[!t]
\centering
{\small
\begin{tabularx}{\textwidth}{@{}l*{5}{>{\raggedleft\arraybackslash}X}@{}}
\toprule
\textbf{Method and provenance} & \multicolumn{1}{c}{\textbf{Func.}} &
\multicolumn{1}{c}{\textbf{Cat.}} & \multicolumn{1}{c}{\textbf{Dom.}} &
\multicolumn{1}{c}{\textbf{Desc.}} & \multicolumn{1}{c}{\textbf{Avg.}} \\
\midrule
\rowcolor{blkTool}\multicolumn{6}{@{}l@{}}{\textit{PFUA-published tool-powered reasoning (prior work)}} \\
\rowcolor{blkTool}DeepSeek-R1$^\dagger$ (671B) & 59.71/38.31 & 47.95/34.15 & 48.36/34.36 & 58.11/35.70 & 53.53/35.63 \\
\rowcolor{blkTool}Kimi-K2-Thinking$^\ddagger$ (1T) & 57.98/35.73 & 67.68/46.90 & 39.76/28.56 & 60.08/36.78 & 56.38/36.99 \\
\rowcolor{blkTool}Qwen3-Max$^\spadesuit$ (preview) & 66.43/44.29 & 72.32/54.18 & 54.98/44.26 & 63.60/41.90 & 64.33/46.16 \\
\midrule
\rowcolor{rowRepro}\multicolumn{6}{@{}l@{}}{\textit{Frontier APIs evaluated by us (reproductions / extensions)}} \\
\rowcolor{rowRepro}DeepSeek-R1 (671B) & 59.57/37.49 & 53.09/37.10 & 46.33/31.89 & 57.61/35.48 & 54.15/35.49 \\
\rowcolor{rowRepro}Kimi-K2-Thinking (1T) & 57.55/35.25 & 67.20/46.30 & 39.40/28.10 & 59.70/36.45 & 56.15/36.40 \\
\rowcolor{rowRepro}Qwen3-Max & 59.66/37.21 & 64.87/46.60 & 45.96/31.99 & 56.95/34.91 & 56.86/37.68 \\
\rowcolor{rowRepro}Claude-Opus-4-8 & 62.70/39.52 & 64.86/47.60 & 43.71/30.51 & 60.43/39.38 & 57.92/39.25 \\
\rowcolor{rowRepro}Claude-Opus-4-7 & 62.03/41.02 & 65.32/47.56 & 42.75/28.58 & 61.01/36.41 & 57.78/38.39 \\
\rowcolor{rowRepro}GPT-5.5 & 65.50/46.01 & 67.89/52.27 & 50.36/37.69 & 61.90/42.65 & \textbf{61.41/44.66} \\
\midrule
\rowcolor{blkSFT}\multicolumn{6}{@{}l@{}}{\textit{Qwen3-8B checkpoints trained and evaluated by us}} \\
\rowcolor{blkSFT}Qwen3-8B (prompted base) & 59.17/36.50 & 58.63/42.15 & 46.72/33.18 & 49.96/29.55 & 53.62/35.33 \\
\rowcolor{blkSFT}Qwen3-8B (SFT) & 59.37/38.28 & 63.95/47.20 & 43.99/33.04 & 50.59/32.22 & 54.47/37.68 \\
\rowcolor{rowOurs}ProtAgent-8B (PFUA SFT+RL) & 62.19/40.85 & 66.91/49.85 & 52.93/39.60 & 58.59/36.16 & \underline{60.15/41.62} \\
\bottomrule
\end{tabularx}}
\caption{Complete PFUA-compatible tool-powered comparison on the Mol-Instructions benchmark.
Each cell reports ROUGE-1 / ROUGE-L recall (\%). Boldface and underline rank only rows evaluated
by us; published PFUA rows are prior-work context and are not ranked against them.
The middle band contains our evaluations of served frontier APIs, not models trained in this
work. Only the bottom-band SFT and SFT+RL checkpoints are post-trained by us.
$^\dagger$DeepSeek-Reasoner~\citep{deepseekr1};
$^\ddagger$Kimi-K2-Thinking~\citep{team2025kimi};
$^\spadesuit$Qwen3-Max-Preview, as reported by PFUA~\citep{fan2026pfua}.}
\label{tab:pfua-complete}
\end{table*}

\begin{table*}[!t]
\centering
{\small
\begin{tabularx}{\textwidth}{@{}l*{5}{>{\raggedleft\arraybackslash}X}@{}}
\toprule
\textbf{Method} & \multicolumn{1}{c}{\textbf{Func.}} & \multicolumn{1}{c}{\textbf{Cat.}} & \multicolumn{1}{c}{\textbf{Dom.}} & \multicolumn{1}{c}{\textbf{Desc.}} & \multicolumn{1}{c}{\textbf{Avg.}} \\
\midrule
\rowcolor{blkSFT}\multicolumn{6}{@{}l@{}}{\textit{Supervised Fine-tuning}} \\
\rowcolor{blkSFT}BioMedGPT (10B) & 5.98/4.28 & 7.97/6.28 & 1.81/1.81 & 4.84/4.12 & 5.15/4.12 \\
\rowcolor{blkSFT}ProtT3 (1.6B) & 15.46/10.81 & 17.36/12.41 & 16.83/12.22 & 19.49/15.96 & 17.28/12.85 \\
\rowcolor{blkSFT}Prot2Text (0.9B) & 16.56/11.61 & 18.24/13.05 & 11.49/9.37 & 49.14/47.38 & 23.86/20.35 \\
\rowcolor{blkSFT}Qwen2.5-3B-SFT & 40.74/30.85 & 41.45/34.52 & 42.60/32.13 & 33.95/25.68 & 39.69/30.79 \\
\midrule
\rowcolor{blkText}\multicolumn{6}{@{}l@{}}{\textit{Text-based Reasoning}} \\
\rowcolor{blkText}BioMedGPT-R1 (17B) & 35.16/26.80 & 27.64/22.22 & 30.60/23.33 & 27.78/20.82 & 30.30/23.29 \\
\rowcolor{blkText}Qwen2.5-3B-R1 & 49.10/38.04 & 61.33/46.54 & 51.01/41.10 & 42.56/32.37 & 51.00/39.51 \\
\midrule
\rowcolor{blkOnline}\multicolumn{6}{@{}l@{}}{\textit{Online LLM Baseline}} \\
\rowcolor{blkOnline}DeepSeek-R1$^\dagger$ (671B) & 26.65/19.68 & 21.86/18.02 & 26.09/19.70 & 16.53/10.76 & 22.78/17.04 \\
\rowcolor{blkOnline}Kimi-K2-Thinking$^\ddagger$ (1T) & 25.08/17.32 & 22.99/17.97 & 27.28/20.07 & 22.30/14.80 & 24.41/17.54 \\
\rowcolor{blkOnline}Qwen3-Max$^\spadesuit$ & 18.53/12.62 & 15.93/12.27 & 15.13/11.28 & 18.59/11.94 & 17.05/12.03 \\
\midrule
\rowcolor{blkRAG}\multicolumn{6}{@{}l@{}}{\textit{Multi-Source RAG}} \\
\rowcolor{blkRAG}DeepSeek-R1$^\dagger$ (671B) & 38.23/25.68 & 32.36/23.75 & 30.35/21.46 & 28.54/16.73 & 32.37/21.91 \\
\rowcolor{blkRAG}Kimi-K2-Thinking$^\ddagger$ (1T) & 22.30/14.14 & 18.56/13.15 & 21.64/17.12 & 45.41/26.44 & 26.98/17.71 \\
\rowcolor{blkRAG}Qwen3-Max$^\spadesuit$ & 37.42/24.52 & 50.66/40.67 & 16.44/12.44 & 42.61/26.34 & 36.78/25.99 \\
\bottomrule
\end{tabularx}}
\caption{PFUA-published families omitted from the main table: supervised fine-tuning,
text reasoning, online models, and multi-source retrieval~\citep{fan2026pfua}. Cells report
ROUGE-1 / ROUGE-L recall (\%)~\citep{rouge}. Systems include BioMedGPT~\citep{biomedgpt},
ProtT3~\citep{prott3}, Prot2Text~\citep{prot2text}, and Qwen2.5 variants~\citep{hui2024qwen2}.
Parentheses give parameter counts; $^\dagger$DeepSeek-Reasoner~\citep{deepseekr1},
$^\ddagger$Kimi-K2-Thinking~\citep{team2025kimi}, and
$^\spadesuit$Qwen3-Max-Preview, as reported by PFUA~\citep{fan2026pfua}.}
\label{tab:pfua-smallscale}
\end{table*}

Our reproduction uses the released 100-sample subset (25 examples per task), evaluation code,
and official PFUA scorer. The task and metric are therefore the same, but the model endpoints
and run times differ from the published experiments. The DeepSeek reproduction has
100/100 sample coverage, uses native tools without heuristic fallback, and passes
41/41 official metric fixtures. Its Func.~and Desc.~scores are within 1.0 ROUGE-1 of the
published values. This agreement is a consistency check, not proof of pipeline equivalence.

Cat.~and Dom.~differ from the published values by +5.14 and $-$2.03 ROUGE-1. With only 25
examples per task, one changed output can move a task mean by about 2 points. Sampling and
endpoint changes may explain these differences, but one run is not enough to separate those
effects from a systematic implementation difference. PFUA also reports one run without variance
estimates or confidence intervals.

\paragraph{Unreproduced published row.}
The published Qwen tool-powered row (64.33/46.16) is the one entry our evaluation does not
reproduce. With the same pipeline, tools, prompt, subset, and scorer, Qwen3-Max reaches
56.86/37.68, a gap of 7.47/8.48 points. This is much larger than the average differences for
DeepSeek ($+0.62$) and Kimi-K2-Thinking ($-0.23$). The published result used a preview endpoint,
whereas our run uses the currently served model; both results are also based on only 25 examples
per task. We therefore keep the published value as prior-work context and compare systems within
our evaluation using only the reproduced rows.

\paragraph{Published families outside active tool use.}
Table~\ref{tab:pfua-smallscale} reports the PFUA results omitted from the compact main-text table.
These values come from PFUA and were not reproduced by us. The SFT and text-reasoning rows are
smaller trained systems. The online and retrieval rows use the same frontier backbones with no
external evidence or passive retrieval, respectively~\citep{rag}. For all three frontier models,
PFUA reports higher ROUGE-1 from online prompting to retrieval and then active tool use
(DeepSeek-R1 22.78 / 32.37 / 53.53, Kimi-K2-Thinking 24.41 / 26.98 / 56.38, and Qwen3-Max
17.05 / 36.78 / 64.33). This prior result motivates our focus on learning a tool-use policy.

\section{Training Configuration}
\label{app:training-config}
Table~\ref{tab:hparams} lists the reinforcement-learning and supervised fine-tuning
hyperparameters for the main-text recipe, including optimization, batching, schedule, and systems.

\paragraph{SFT corpus.}
The corpus contains 987 training and 64 validation trajectories. The training split comprises
567 action examples, 120 evidence-to-answer examples, and 300 multi-turn trajectories.
Admission requires valid tool-call
format, a substantive evidence-bearing answer, and task-critical tool coverage.

\begin{table*}[t]
\centering{\small
\begin{minipage}[t]{0.47\textwidth}
\centering
\textbf{(a) Reinforcement learning}\\
\setlength{\tabcolsep}{4pt}%
\begin{tabular}{@{}ll@{}}
\toprule
Hyperparameter & Value \\
\midrule
\multicolumn{2}{@{}l}{\textit{Objective and optimization}}\\
Advantage estimator & Batched Step-Level REINF. \\
Policy-gradient variant & Vanilla (no GRPO) \\
Return normalization & Identity \\
Advantage whitening & Enabled \\
Advantage clip & 0.2 \\
Policy epochs & 1 \\
KL coefficient $\beta$ & 0.001 \\
Discount $\gamma$ & 1.0 \\
Entropy coefficient & 0 \\
Max gradient norm & 1.0 \\
Learning-rate schedule & Constant within each stage \\
Stage learning rates & $5{\times}10^{-7}/5{\times}10^{-8}/5{\times}10^{-8}$ \\
Weight decay & 0 \\
\midrule
\multicolumn{2}{@{}l}{\textit{Batching and schedule}}\\
Trajectories/prompt $K$ & 4 \\
Prompt instances/update & 8 \\
Interaction steps/episode & 4 \\
Updates per stage & 50 (ckpt every 10) \\
Sequence length & 4{,}096 \\
Tokens per response & 1{,}024 \\
\midrule
\multicolumn{2}{@{}l}{\textit{Systems}}\\
Precision & bfloat16 \\
Training backend & FSDP2 \\
Inference engine & vLLM (TP\,2) \\
Chat template & Qwen2.5 \\
\bottomrule
\end{tabular}
\end{minipage}
\hfill
\begin{minipage}[t]{0.47\textwidth}
\centering
\textbf{(b) Supervised fine-tuning}\\
\setlength{\tabcolsep}{4pt}%
\begin{tabular}{@{}ll@{}}
\toprule
Hyperparameter & Value \\
\midrule
Base model & Qwen3-8B-Instruct \\
Epochs & 1 \\
Learning rate & $1\times10^{-6}$, cosine \\
Weight decay & 0.01 \\
Batch $\times$ grad-accum & $1\times1$ \\
Sequence length & 8{,}192 \\
Label shifting & Enabled \\
Precision & bfloat16 \\
Training backend & Megatron \\
Parallelism & TP2, PP2, CP2, seq.-par. \\
Chat template & Qwen2.5 \\
Train / val examples & 987 / 64 \\
\bottomrule
\end{tabular}
\end{minipage}}
\caption{Training hyperparameters: (a)~reinforcement learning and (b)~supervised fine-tuning.
The reinforcement-learning schedule uses three 50-update stages. Each restart uses the previous
checkpoint, a fresh optimizer, and a new reference anchor.}
\label{tab:hparams}
\end{table*}

\section{Tool Catalog}
\label{app:tool-catalog}
Table~\ref{tab:tool-catalog} lists the enabled tools, backends, and dependencies in the project
registry. Paper protocols expose the task-relevant tools through their prompts and tool
interfaces.

\begin{table*}[!tbp]
\centering{\small
\setlength{\tabcolsep}{5pt}%
\begin{tabular}{llll}
\toprule
Tool & Category & Backend & Database / dependency \\
\midrule
\multicolumn{4}{@{}l}{\textit{Native / heuristic computation}}\\
\texttt{seq\_basic\_props} & Sequence properties & Native (heuristic) & None \\
\texttt{mmseqs2\_besthit\_uniprot} & Homology search & Native (MMseqs2) & Swiss-Prot \\
\texttt{pfam\_hmmscan} & Domain detection & Native (HMMER) & Pfam-A \\
\texttt{tmbed\_predict} & Membrane topology & Native (TMbed) & TMbed weights \\
\texttt{signalp\_predict} & Signal peptide & Native (SignalP\,6.0) / heuristic & SignalP\,6.0 weights \\
\texttt{disprot\_region} & Disordered regions & Native (IUPred3) / heuristic & IUPred3 \\
\texttt{ec\_number\_predict} & Enzyme class (EC) & Native (CLEAN) / heuristic & CLEAN weights \\
\texttt{protein\_design\_generate} & Sequence design & Heuristic & None \\
\texttt{design\_gate\_verify} & Design gating & Heuristic & None \\
\midrule
\multicolumn{4}{@{}l}{\textit{Skill-routed API lookup}}\\
\texttt{string\_db\_lookup} & Protein interactions & API & STRING \\
\texttt{quickgo\_lookup} & GO annotation & API & QuickGO \\
\texttt{reactome\_lookup} & Pathway membership & API & Reactome \\
\bottomrule
\end{tabular}}
\caption{Enabled tool catalog. Entries marked ``/ heuristic'' have an alternative implementation
in the registry. Paper-reported evaluations select native mode, which raises an error rather
than silently falling back when a required native backend is unavailable. Reproduction requires
the tool and database versions used in the corresponding run. Native backends and reference resources are
MMseqs2~\citep{mmseqs2} over Swiss-Prot~\citep{uniprot}, HMMER~\citep{hmmer} over
Pfam-A~\citep{pfam}, TMbed~\citep{tmbed}, SignalP\,6.0~\citep{signalp6},
IUPred3~\citep{iupred3}, CLEAN~\citep{clean}, and Gene Ontology terms~\citep{geneontology}
retrieved through QuickGO.}
\label{tab:tool-catalog}
\end{table*}

\section{Prompt Templates}
\label{app:prompts}
During training-data construction, each example is assigned one of four system-prompt variants
according to Table~\ref{tab:prompt-variants}. The variants provide different amounts of
reasoning and tool-routing guidance while preserving the function-calling interface and the
\texttt{<answer>{\ldots}</answer>} output contract. Evaluation uses one frozen concise prompt
rather than sampling the training variants. Only PFUA-full explicitly requires the model to
state a hypothesis before a tool call and update it after the observation; the other variants
do not impose that rule.

\begin{center}
{\footnotesize
\setlength{\tabcolsep}{3pt}%
\begin{tabularx}{\columnwidth}{@{}lccY@{}}
\toprule
Variant & Weight & Tokens & Contents \\
\midrule
PFUA-lite & 0.40 & 200--300 & Role / task / tools / output skeleton \\
Generic+ & 0.30 & $\sim$120 & Four-line prompt + answer format \\
PFUA-full & 0.15 & $\sim$625 & Full structured system prompt \\
Skill-aware & 0.15 & $\sim$150 & Per-task tool routing + concise output \\
\bottomrule
\end{tabularx}}
\refstepcounter{table}\label{tab:prompt-variants}
\smallskip
{\small\textbf{Table \thetable:} Prompt variants sampled during training-data construction. Weights are assignment
probabilities and sum to one; token counts are approximate under the model tokenizer.}
\end{center}

\section{Representative Evidence-Conversion Trace}
\label{app:trajectory}
Figure~\ref{fig:trajectory} shows a condensed rendering of a held-out catalytic-activity
trajectory. A remote
homology hit proposes a bifunctional NAD(P)H-hydrate repair enzyme, while an independent Pfam scan
recovers the complementary N-terminal and dehydratase domains. The policy must select the
supported activities and express both reactions in the requested form. Internal run and
checkpoint identifiers and long metadata fields are omitted.

\paragraph{What the trace illustrates.}
The example shows evidence retrieval followed by answer construction. The homology search proposes
a candidate function, and the domain scan checks whether the query has the two modules expected
for the bifunctional enzyme. The final answer combines these results and gives the requested
reactions. This illustrates the workflow, but a single trace does not isolate the contribution
of reasoning beyond retrieval or establish how often the model uses evidence correctly.

\begin{figure*}[!t]
\centering
\begin{tcblisting}{
  enhanced,
  width=\textwidth,
  colback=trajBg,
  colframe=trajFrame,
  arc=1.5mm,
  boxrule=0.5pt,
  left=5pt, right=5pt, top=5pt, bottom=5pt,
  listing only,
  listing options={
    basicstyle=\small\ttfamily,
    escapeinside=@@,
    breaklines=true,
    breakatwhitespace=false,
  },
}
@\textcolor{trajSys}{\textbf{[system]}}@ You are a protein analysis agent with access to bioinformatics tools. Call tools one at a time; then wrap the final answer in <answer>...</answer>. Do not hallucinate tool results---use only information from actual outputs.

@\textcolor{trajUsr}{\textbf{[user]}}@ Analyze the provided 508-aa protein sequence and describe the enzyme's catalytic activity, including the chemical reactions it facilitates.

@\textcolor{trajAsst}{\textbf{[assistant, step 1]}}@ I will first identify a functionally informative homolog and then verify whether the query carries the expected domain architecture.
  <tool_call>{"name":"mmseqs2_besthit_uniprot", "arguments":{"sequence_ref":"query"}}

@\textcolor{trajTool}{\textbf{[tool]}}@ Best hit: Q2RGI2, 37.2% identity over 518 aligned residues, e-value 2.7e-83. Annotation: bifunctional NAD(P)H-hydrate repair enzyme Nnr; EC 4.2.1.136 and EC 5.1.99.6. Reported reactions include ADP-dependent dehydration of (6S)-NAD(P)HX and epimerization of (6R)- to (6S)-NAD(P)HX.

@\textcolor{trajAsst}{\textbf{[assistant, step 2]}}@ The hit is remote rather than an identical record. I will test whether both functional modules expected for Nnr are present in the query.
  <tool_call>{"name":"pfam_hmmscan", "arguments":{"sequence_ref":"query"}}

@\textcolor{trajTool}{\textbf{[tool]}}@ Pfam: YjeF_N (PF03853; residues 27--195; e-value 3.2e-44) and Carb_kinase (PF01256; residues 253--508; e-value 3.8e-56), annotated as an NAD(P)H-hydrate dehydratase domain.

@\textcolor{trajAsst}{\textbf{[assistant, final; task-relevant excerpt]}}@ <answer>The query is supported as a bifunctional NAD(P)H-hydrate repair enzyme. Its dehydratase activity (EC 4.2.1.136) converts (6S)-NADHX + ADP to NADH + AMP + phosphate + H(+), with the analogous NADPHX reaction producing NADPH. Its epimerase activity (EC 5.1.99.6) interconverts the (6R)- and (6S)-NAD(P)HX forms, allowing both damaged epimers to enter the repair pathway. The YjeF_N and Carb_kinase domain pair independently supports this bifunctional assignment.</answer>
\end{tcblisting}
\caption{Condensed held-out evidence-conversion trace (catalytic-activity task). A 37.2\%-identity
homology hit proposes the function, Pfam independently recovers both expected modules, and the
task-relevant answer excerpt selects and reformulates the supported catalytic activities. Long
sequence, metadata, and answer fields outside the requested catalytic activity are omitted for
readability. The code-and-data supplement does not include the complete raw trajectory.}
\label{fig:trajectory}
\end{figure*}

\section{Code, Data, and Evaluation Checks}
\label{app:reproducibility}
The code-and-data supplement contains benchmark inputs, scoring code, tool interfaces, reward
implementations, configuration summaries, validation scripts, and tests. It does not include
model checkpoints, raw model outputs, result summaries, or supplementary-only ablation artifacts.
It supports running and scoring new evaluations, but is not a complete archive from which every
reported table can be recomputed.

\paragraph{Records needed for reproduction.}
A reproducible run requires a protocol manifest identifying the benchmark, checkpoint, prompt,
decoding settings, code revision, and tool and database versions. Each sample record should retain
the tool calls, observations, final answer, and termination state. These records, together with
per-sample scores, are needed to verify coverage and recompute paired comparisons.

\paragraph{Validation tools.}
The supplied coverage validator checks for missing, unknown, or duplicate sample IDs; the scorer
keeps unanswered episodes in the denominator. Tool-integrity checks look for known fallback
signatures, while native-mode preflight checks required dependencies. These tools help validate
new runs; their inclusion is not a substitute for the records of the reported experiments.

\end{document}